# Controlled spontaneous emission

Jae-Seung Lee,[1] Mary A. Rohrdanz,[2] and A. K. Khitrin[1]

[1] *Department of Chemistry, Kent State University, Kent, Ohio 44242-0001*

[2] *Department of Chemistry, Walsh University, North Canton, Ohio 44720-3336*

**Abstract**

The problem of spontaneous emission is studied by a direct computer simulation of the dynamics of a combined system: atom + radiation field. The parameters of the discrete finite model, including up to 20k field oscillators, have been optimized by a comparison with the exact solution for the case when the oscillators have equidistant frequencies and equal coupling constants. Simulation of the effect of multi-pulse sequence of phase kicks and emission by a pair of atoms shows that both the frequency and the linewidth of the emitted spectrum could be controlled.

## 1. INTRODUCTION

The process of spontaneous light emission is a collective quantum dynamics of an atom and electromagnetic radiation field. Since the solution has been proposed by Dirac [1] and Fermi [2], many authors revisited this problem with analytical methods or computer simulations in order to reveal more details of this process. Today, greatly enhanced computational capabilities allow a direct simulation of spontaneous emission not only in its classical formulation, but also make possible simulations of more complex dynamics, for example, emission by a multi-atom system, or under perturbation of an atom by a train of laser pulses.

In this paper, we will follow the concepts of the Fermi's work [2], i.e. it will be assumed that the atom is placed in a box and coupled to the normal modes (quantum oscillators) of the electromagnetic radiation field inside the box. The goal is to find the limiting behavior when the size of the box goes to infinity. In the initial state, the atom is in its first excited state and the field oscillators are in their ground states (the Weisskopf – Wigner model [3], the Jaynes-Cummings model [4]). Further, if one neglects multi-photon processes, the evolution can be restricted to a small subspace of the entire Hilbert space, where only the ground and the first excited state of each field oscillator are included. The basis set that spans this subspace is $\{|\Psi_k\rangle\}$. $|\Psi_0\rangle$ corresponds to the state with the atom in its excited state and all field oscillators in their ground state. $|\Psi_k\rangle$ with $k \neq 0$ corresponds to the state in which the $k$-th oscillator is in its first excited state, while the atom and other oscillators are in their ground states. This approximation is based on the notion that when the size of the box increases to infinity, the coupling constants between the atom and field oscillators decrease to zero and, therefore, only near-resonance oscillators are important. As a result, in this "single-photon subspace" both the atom and the field oscillators are represented by the two-level systems. The Hamiltonian is

$$H = \sum_k E_k S_k^{\ z} + \sum_{k \neq 0} \eta_k (S_0^{\ +} S_k^{\ -} + S_0^{\ -} S_k^{\ +}) \, , \tag{1}$$

where $E_k$ is the energy difference between the excited and ground states of the atom ($k = 0$) and the oscillators ($k \neq 0$), $\eta_k$ is the coupling constant between the atom and the $k$-th field oscillator, $S^{\pm} = S^x \pm iS^y$, $S^{\alpha} = (1/2)\sigma^{\alpha}$, and $\sigma^{\alpha}$, $\alpha$ = x,y,z, are the Pauli matrices. In the interaction frame, obtained by the transformation $U = \exp(-it\,\hbar^{-1} E_0 \sum_k S_k^{\ z})$, the Hamiltonian is

$$H = \sum_k \varepsilon_k S_k^{\ z} + \sum_{k \neq 0} \eta_k (S_0^{\ +} S_k^{\ -} + S_0^{\ -} S_k^{\ +}) \, , \tag{2}$$

where $\varepsilon_k = E_k - E_0$ is the resonance offset of the $k$-th oscillator.

In the $\{|\Psi_k\rangle\}$ basis the Hamiltonian (2) is represented by the matrix

$$H = \begin{pmatrix} & \varepsilon_2 & & \eta_2 & & \\ & & \varepsilon_1 & \eta_1 & & \\ \ldots \eta_2 & & \eta_1 & 0 & \eta_{-1} & \eta_{-2} \ldots \\ & & & \eta_{-1} & \varepsilon_{-1} & \\ & & & \eta_{-2} & & \varepsilon_{-2} \end{pmatrix} , \tag{3}$$

where only the non-zero elements of the matrix are shown. The wavefunction at any moment in time is given by

$$\Psi(t) = \exp(-it\hbar^{-1} H)\ \Psi(0) = \sum_k a_k(t)\ \Psi_k \, . \tag{4}$$

In the initial state $a_0 = 1$, $a_k = 0$ ($k \neq 0$). $|a_0(t)|^2$ gives the probability of the atom to be in the excited state, while the probabilities of the $k$-th oscillator to be excited, $|a_k(t)|^2$, describe the spectrum of the emitted light. The evolution (4) is calculated in this work by a direct diagonalization of the Hamiltonian (3).

The paper is organized as follows. Section 2 provides calculations on the $2p_z \rightarrow 1s$ transition of a hydrogen atom in a 3-dimensional box. Section 3 details a computationally efficient pseudo-one-dimensional model, which has an analytical solution and can be used to analyze the limit of infinite number of the field oscillators. The results obtained in this section are used to optimize the parameters of a discrete finite model for direct numerical simulation in situations when analytical solutions are not available. In sections 4 and 5 we examine two such problems, namely, a single atom acted upon by a train of composite laser pulses, and spontaneous emission by a system of two coupled atoms. In sections 3, 4, and 5 we consider general two-level atoms, rather than the specific example of the hydrogen transition in section 2. Conclusions are presented in the final section.

Some simulations for an atom in a 3D box have been performed with a cluster of supercomputers of the Ohio Supercomputer Center. The rest of the calculations have been done by using a Dell Precision Workstation equipped with two dual-core 64-bit processors operating at 3.7 GHz.

## 2. 3D BOX

As an example with realistic physical parameters, we present the results of simulation for a hydrogen atom placed in the center of a 3D box. The atom, initially in the $2p_z$ excited state, undergoes a spontaneous transition to the 1s ground state.

For the Hamiltonian (3), numerical simulation becomes impractical due to memory constraints for matrices larger than roughly 20k by 20k. The box size has been optimized to allow a substantial number of the oscillators having frequencies within the linewidth of the emitted spectrum and, at the same time, to make a truncation frequency significantly larger than this linewidth. A small distortion of the box dimensions from cubic has been introduced to reduce degeneracy and make distribution of the oscillator frequencies more uniform.

The electromagnetic radiation field inside the box is written as a superposition of planar standing waves with the electric field proportional to cos $k_x x$ cos $k_y y$ cos $k_z z$. The allowed wave vector components, $k_\alpha$, are chosen to yield a node in the electric field at the box boundaries. The sine waves have been omitted because, in the dipolar approximation used in this work, they are not coupled to the atom, which is located in the center of the box. The frequency of the wave with the wave number $\mathbf{k} = (k_x, k_y, k_z)$ is

$$\omega_k = kc, \quad k = (k_x^2 + k_y^2 + k_z^2)^{1/2}, \tag{5}$$

where $c$ is the speed of light. For each allowed **k** there is a freedom to choose two independent polarization vectors, perpendicular to one another and **k**. It is convenient to choose one of the polarization vectors **ρ** in the (**z**,**k**) plane, where **z** is the unit vector along the $z$-axis. Then, only the wave with this polarization will be coupled to the atom, since the atom has only $z$-component of the transition dipolar moment between the states $2p_z$ and 1s. After the normal modes are defined, as it is described above, quantization is introduced in a conventional fashion [1,2] by assuming that each normal mode is a quantum oscillator. The coupling constants, $\eta_k$, in Hamiltonians (1-3) are [5] (SI units):

$$\eta_k = -\mu\, \mathbf{z}\cdot\boldsymbol{\rho}\, (\hbar\, \omega_k/\varepsilon_0 V)^{1/2} = -\mu \sin(\theta_k)\, (\hbar\, \omega_k/\varepsilon_0 V)^{1/2}, \qquad (6)$$

where μ is the transition dipole moment, $\varepsilon_0$ is the dielectric permeability of vacuum, $V$ is the box volume, and $\theta_k$ is the angle between the corresponding wave vector and the $z$-axis. In our simulations, we used $\mu = e \langle 2p_z|\, z\, |1s\rangle = ea_0 4(2/3)^5 \sqrt{2} = -6.31582621 \times 10^{-30}$ C·m, calculated with the exact eigenfunctions of non-relativistic Hamiltonian for the hydrogen atom, where $e$ is the electronic charge and $a_0$ is the Bohr radius. In $k$-space, wave vectors included in the simulation have been picked from the spherical shell $k_0 - \Delta < k < k_0 + \Delta$, where $k_0 = \omega_0/c$, and $\omega_0$ is the atom's transition frequency. Tables 1 and 2 provide the relevant spectroscopic and simulation parameters. The results of the simulation for $N$ = 20,820 field oscillators are displayed in Figs. 1-3.

**Table 1** Spectroscopic parameters (NIST data [6])

| | |
|---|---|
| Emission wavelength | $\lambda_0$ = 121.566824 nm |
| Emission center frequency | $\nu_0 = \lambda_0/c$ = 2.466 071 32 × $10^{15}$ Hz |
| Einstein A coefficient | A = 6.2648 × $10^8$ $s^{-1}$ |
| Corresponding lifetime | τ = 1/A = 1.5962 ns |

**Table 2** Simulation parameters

| | |
|---|---|
| Number of oscillator states | 20 820 |
| Box length in x-direction | 0.210 259 195 465 634 0 mm |
| Box length in y-direction | 0.210 049 146 319 314 7 mm |
| Box length in z-direction | 0.210 469 665 130 764 7 mm |
| Spectral range of calculation | Δ = 8.186 561 5 $m^{-1}$ |
| Transition dipole moment | $\mu$ = -6.31582621 × $10^{-30}$ C·m |

Fig. 1 shows the decay of the population of the atomic excited state $|a_0(t)|^2$. In the limit $N$, $V \rightarrow \infty$, one would expect an exponential decay, which is consistent with a simple concept of a transition probability per unit time given by the Fermi's "golden rule":

$$1/\tau_F = 2\pi <|\eta_k|^2> \rho_\omega, \quad (7)$$

where $<|\eta_k|^2>$ is the average square of the coupling constant and $\rho_\omega$ is the spectral density of the field oscillators at the transition frequency. The exponential decay shown for comparison in Fig. 1, $\exp(-t/\tau)$, uses $\tau$ from Table 1.

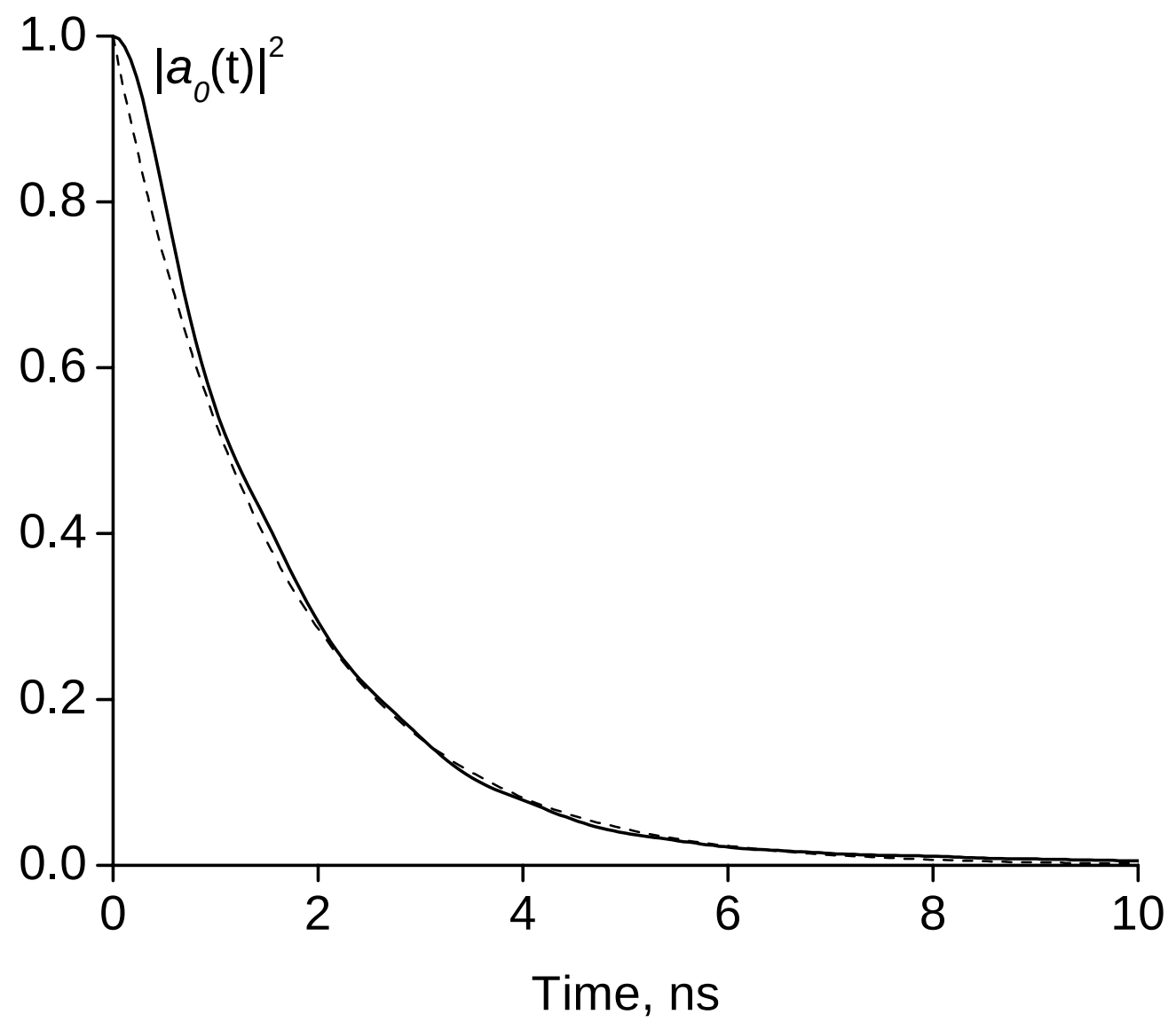

**Fig. 1** Excited-state population of the atom as a function of time. The solid line is the simulation result; the dashed line is $\exp(-t/\tau)$, with $\tau$ from Table 1.

At $t \rightarrow \infty$, the probabilities of finding the field oscillators excited, $|a_k(t)|^2$ ($k \neq 0$), describe the emission spectrum. Fig. 2 shows the oscillator excitation probabilities at $t$ = 17.2 ns, as a function of oscillator frequency. A Lorentzian profile with the full width at half height $1/\tau$ and τ from the Table 1 is shown for comparison. These probabilities depend not only on each oscillator's frequency, but also on the orientation of the oscillator's wave vector with respect to the *z*-axis. The probability is proportional to $\sin^2(\theta_k)$, it is maximal for wave vectors in the *xy*-plane, and approaches zero for wave vectors parallel to the *z*-axis. Fig. 3 displays the angular distribution of the emitted photon; the expected $\sin^2(\theta_k)$ dependence is shown as a boundary. This double dependence, on frequency and orientation, is responsible for the dots that appear below the simulation-data envelopes in Figs. 2 and 3. For example, in Fig.2 near-resonant oscillators without a favorable orientation appear under the envelope. Similarly in Fig.3, oscillators in the *xy*-plane, but far off-resonant, appear under the data envelope.

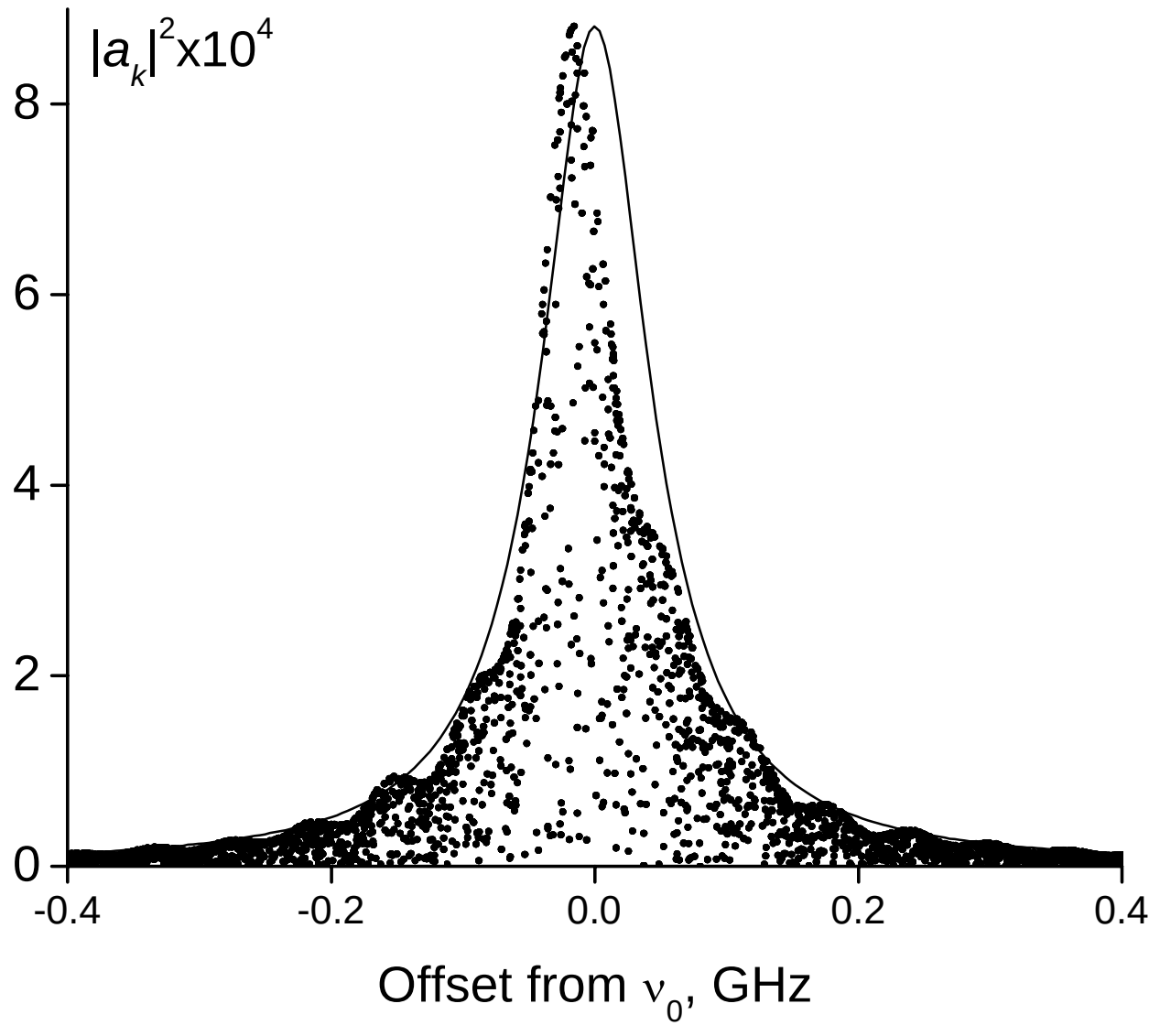

**Fig. 2** Probability of excitation of the field oscillators at time $t$ = 17.2 ns, as a function of each oscillator's frequency. Each dot represents one oscillator in the calculation; the solid line is a Lorenzian lineshape.

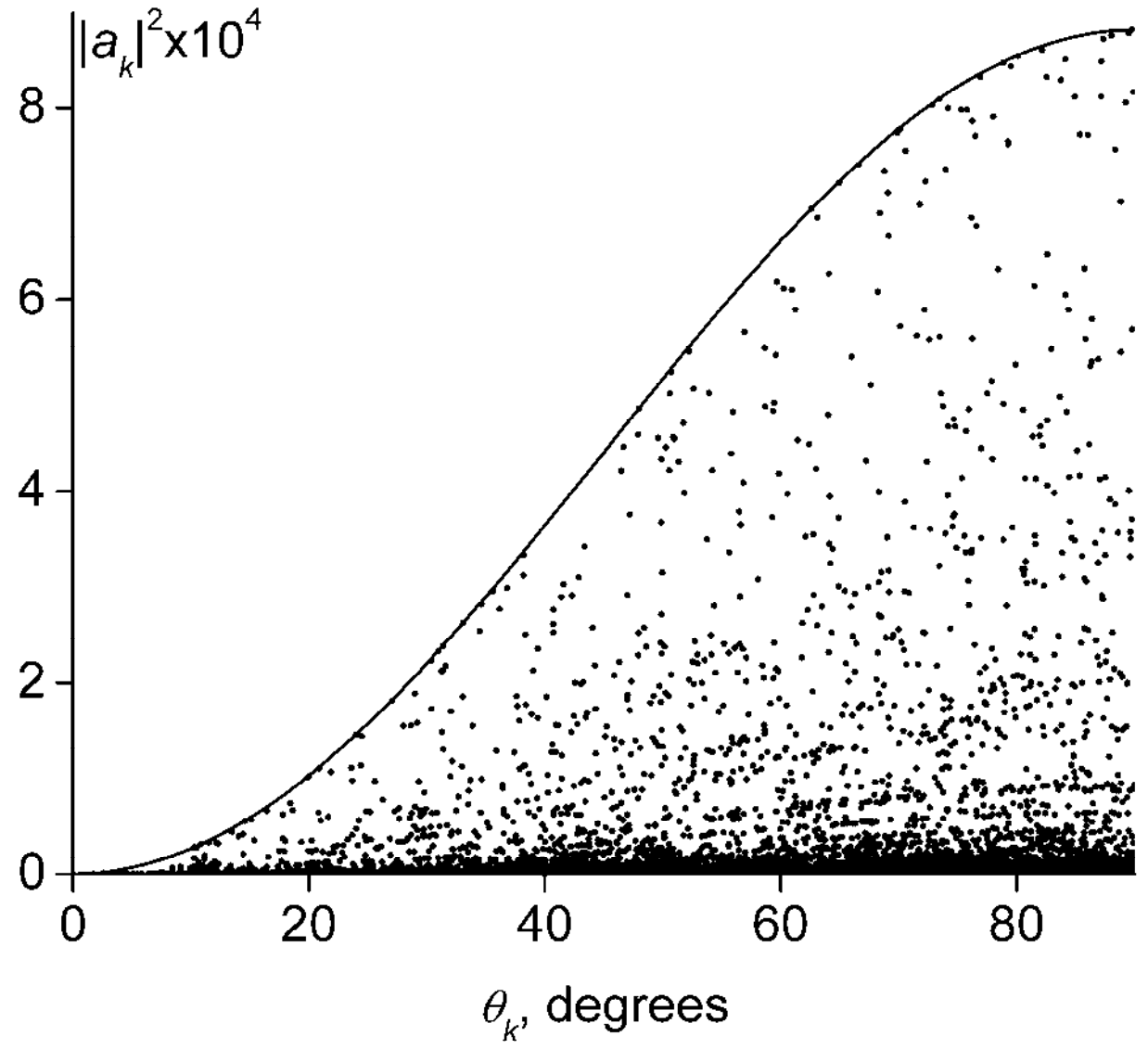

**Fig. 3** Angular distribution of the probability for the emitted photon at $t$ = 17.2 ns. Each dot represents an oscillator in the calculation; the solid line is $\sin^2(\theta_k)$.

As one can see in Figs. 1 and 2, there are noticeable deviations from the behavior expected in the thermodynamic limit even for ~20k field oscillators. Among the reasons that make a discrete 3D model computationally inefficient are variable coupling constants and non-uniform "random" distribution of the oscillators' frequencies. In the limit $V \to \infty$ one expects the dynamics to depend only on average quantities [see Eq. (7)]. Therefore, the "1D model" with equal coupling constants and equidistant oscillator frequencies, discussed in the next section, may better approach the thermodynamic limit for a fixed number of the field oscillators. Another advantage of this model is that for a conventional problem of spontaneous emission there exists an analytical solution, which allows for calculations with even larger values of $N$.

## 3. PSEUDO-1D MODEL: THE EXACT SOLUTION

The simplified form of the Hamiltonian (3) used in this and the following sections has equal coupling constants $\eta_k = \eta$ and equidistant spacing between the frequencies of the oscillators $\varepsilon_k = k\varepsilon$. The goal is to find the behavior at $\eta \to 0$, $\varepsilon \to 0$, $\eta^2/\varepsilon$ = const. The "golden rule" (7) now becomes

$$1/\tau_F = 2\pi\, \eta^2/\varepsilon. \tag{8}$$

With this model, the problem of spontaneous emission can be solved analytically [7-9]. Since we used a different approach, we briefly describe the major steps of the calculation below.

The Hamiltonian can be defined as

$$H\left|\Psi_k\right\rangle = \begin{cases} k\varepsilon\left|\Psi_k\right\rangle + \eta\left|\Psi_0\right\rangle & \text{if } k \neq 0 \\ \eta\sum\limits_{l\neq 0}\left|\Psi_l\right\rangle & \text{if } k = 0 \end{cases} \tag{9}$$

where $k$ spans from $-\infty$ to $+\infty$. Suppose that $\lambda_k$ and $\left|\Phi_k\right\rangle$ are the $k$-th eigenvalue and the corresponding eigenvector of the Hamiltonian, i.e., $H\left|\Phi_k\right\rangle = \lambda_k\left|\Phi_k\right\rangle$. By inserting $\left|\Phi_k\right\rangle = \sum\limits_{l=-\infty}^{\infty}\alpha_k^l\left|\Psi_l\right\rangle$ into the equation $H\left|\Phi_k\right\rangle = \lambda_k\left|\Phi_k\right\rangle$, one obtains the characteristic equations:

$$\eta\sum_{l\neq 0}\alpha_k^l = \lambda_k\alpha_k^0 \tag{10a}$$

$$\varepsilon l\alpha_k^l + \eta\alpha_k^0 = \lambda_k\alpha_k^l \qquad \text{for } l \neq 0. \tag{10b}$$

Therefore, the eigenvalues satisfy the equation

$$\eta^2 \sum_{l \neq 0} \frac{1}{\lambda_k - \varepsilon l} = \lambda_k \,, \tag{11}$$

from which

$$\lambda_0 = 0 \tag{12a}$$

and

$$\tan \frac{\pi \lambda_k}{\varepsilon} = \frac{\pi \eta^2}{\varepsilon} \frac{\lambda_k}{\lambda_k^2 + \eta^2} \qquad \text{for } k \neq 0 \,. \tag{12b}$$

It is obvious that $\lambda_{-k} = -\lambda_k$. From the above equations and the normalization condition, the coefficients $\alpha_k^l$ can be expressed as follows. For $\lambda_0 = 0$,

$$\alpha_0^0 = \sqrt{\frac{3}{3 + (\pi\eta/\varepsilon)^2}} \tag{13a}$$

and

$$\alpha_0^l = -\frac{\eta}{l\varepsilon} \alpha_0^0 = -\frac{\eta}{l\varepsilon} \sqrt{\frac{3}{3 + (\pi\eta/\varepsilon)^2}} \qquad \text{for } l \neq 0 \,, \tag{13b}$$

where $\alpha_0^0$ is chosen to be real. Similarly, for the other eigenvalues,

$$\alpha_k^0 = \sqrt{\frac{1}{3 + (\pi\eta/\varepsilon)^2 + (\pi\lambda_k/\eta)^2}} \tag{14a}$$

and

$$\alpha_k^l = \frac{\eta}{\lambda_k - \varepsilon l} \alpha_k^0 = \frac{\eta}{\lambda_k - \varepsilon l} \sqrt{\frac{1}{3 + (\pi\eta/\varepsilon)^2 + (\pi\lambda_k/\eta)^2}} \qquad \text{for } l \neq 0 \,. \tag{14b}$$

Now, it is straightforward to calculate the probability amplitude that the system is in the state $|\Psi_k\rangle$. Since the initial state is $|\Psi_0\rangle$, the probability amplitude is given by

$$\left\langle \Psi_k \left| \exp\left(-iHt\right) \right| \Psi_0 \right\rangle = \sum_l \alpha_l^k (\alpha_l^0)^* \exp(-i\lambda_l t) . \tag{15}$$

For $k = 0$,

$$\begin{aligned} \left\langle \Psi_0 \left| \exp\left(-iHt\right) \right| \Psi_0 \right\rangle &= \left|\alpha_0^0\right|^2 \exp(-i\lambda_0 t) + \sum_{l\neq 0} \left|\alpha_l^0\right|^2 \exp(-i\lambda_l t) \\ &= \frac{3}{3 + \left(\pi\eta/\varepsilon\right)^2} + \sum_{l\neq 0} \frac{\exp\left(-i\lambda_l t\right)}{3 + \left(\pi\eta/\varepsilon\right)^2 + \left(\lambda_l/\eta\right)^2} \end{aligned} . \tag{16}$$

The absolute square of this quantity gives the probability that the atom stays in the excited state. According to the Fermi's "golden rule" (8), this probability or population decays exponentially with the characteristic time $\varepsilon(2\pi\eta^2)^{-1}$. Fig. 4 shows the difference between the population predicted by the "golden rule" and the probability obtained from Eq. (16) when the summation in (16) is truncated at $|l|>500000$ for several values of $\eta/\varepsilon$. If the number of terms kept in the summation is not sufficiently large, a damping oscillation in the population decay curve appears at short times, although its amplitude is smaller than the initial difference resulting from the truncation. One million terms is sufficient to remove the truncation effects, as seen in Fig 4, where the initial differences are less than 0.001 and the damping oscillations are almost absent. Instead, Fig. 4 shows that two curves are different in the middle of the decaying process but that the difference diminishes as the ratio $\eta/\varepsilon$ increases. Therefore, in the limit of large $\eta/\varepsilon$ and $N \to \infty$ the probability that the atom stays in the excited state follows the exponentially decaying curve predicted by the "golden rule". The proof of this has been provided earlier in Refs. [7,8].

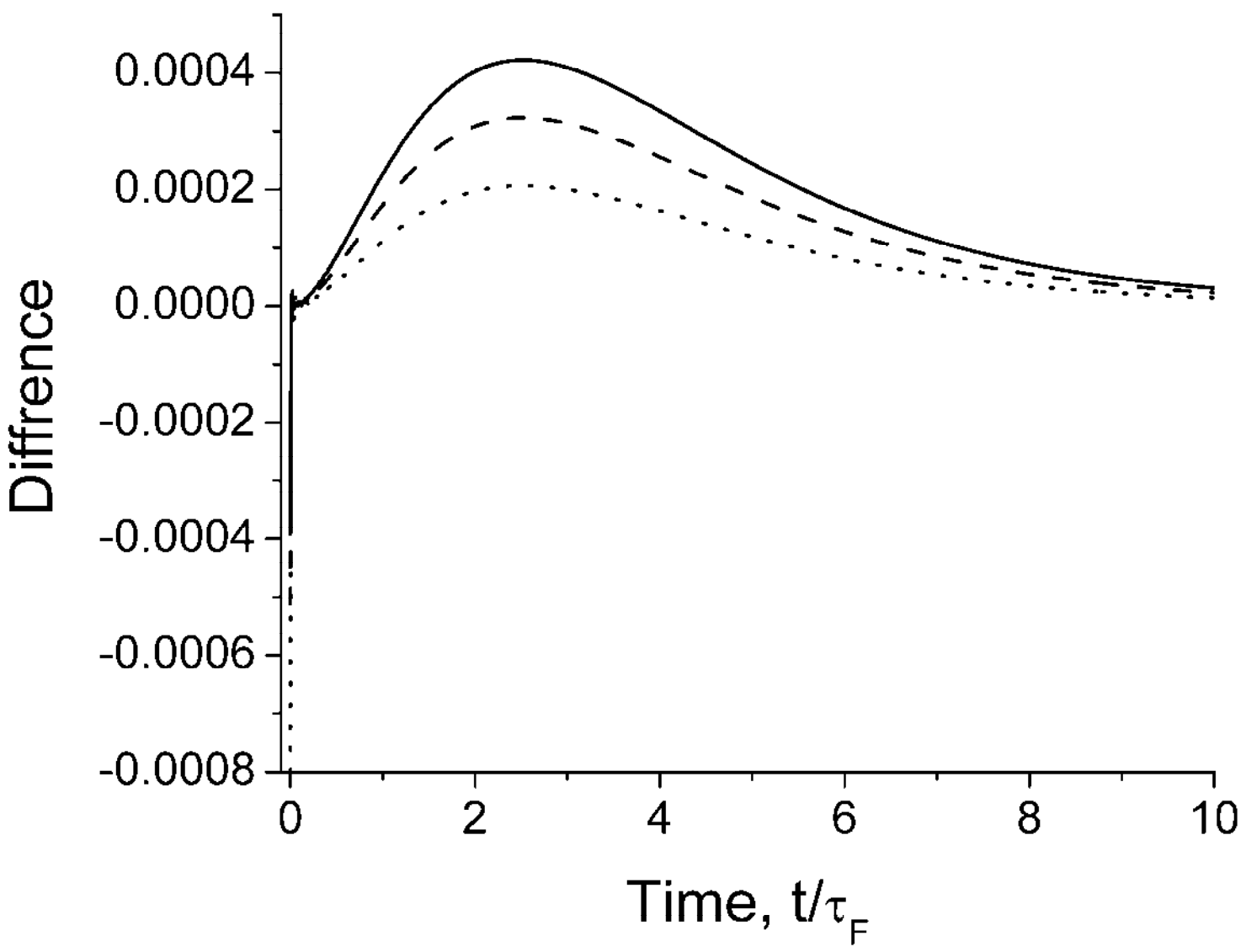

**Fig. 4** Difference between the population predicted by the golden rule and the probability evaluated from Eq. (16) with ($10^6$+1) states included. There exist initial differences of -0.0004, -0.0005, and -0.0008 for $\eta/\varepsilon = 7$ (solid), 8 (dash), and 10 (dotted), respectively, due to the truncation in summation.

For $k \neq 0$,

$$\begin{aligned}\langle \Psi_k | \exp(-iHt) | \Psi_0 \rangle &= \sum_l \frac{\eta}{\lambda_l - \varepsilon k} \left|\alpha_l^0\right|^2 \exp(-i\lambda_l t) \\ &= -\frac{\eta}{\varepsilon k} \frac{3}{3 + (\pi\eta/\varepsilon)^2} + \sum_{l \neq 0} \frac{\eta}{\lambda_l - \varepsilon k} \frac{\exp(-i\lambda_l t)}{3 + (\pi\eta/\varepsilon)^2 + (\lambda_l/\eta)^2}\end{aligned} \quad . \tag{17}$$

Eq. (17) allows evaluating the probability distribution of the states after the spontaneous emission by the atom. With $\varepsilon = 10^{-10}$, $\eta/\varepsilon = 10$, $t = 1.6 \times \varepsilon/\eta^2$, and ($10^6$+1) states, the probability distribution perfectly fits the Lorentzian curve with the full width at half height $2\pi\eta^2/\varepsilon$ (not shown).

In direct numerical simulations using modern computers, one is limited by about 20k field oscillators. In this case, the choice of the ratio $\eta/\varepsilon$ becomes important. Larger values of $\eta/\varepsilon$ improve the behavior at intermediate times but spoil the short-time dynamics. A compromise value of $\eta/\varepsilon$ is needed to minimize the error in the entire time range. As one can see in Fig. 5, for 15 k oscillators the optimal value of $\eta/\varepsilon = 2.4$ makes the error in the population decay curve below 0.0035 at all times.

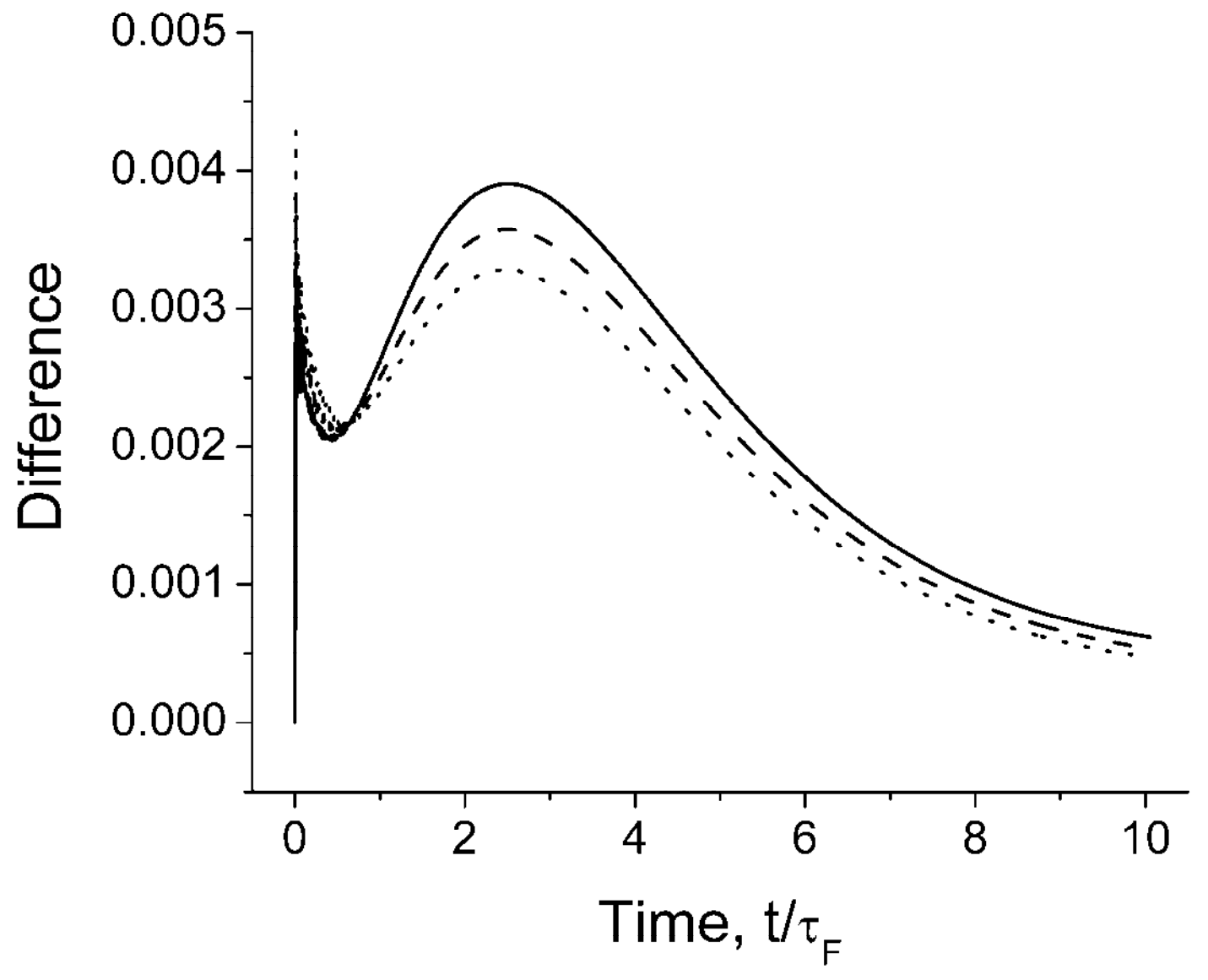

**Fig. 5** Deviations from the exponential population decay for 15 k field oscillators and $\eta/\varepsilon$ = 2.3 (solid), 2.4 (dash), and 2.5 (dots).

### 4. MULTI-PULSE TRAIN OF PHASE KICKS

The interaction between the atom and an individual field oscillator [see Eq. (1)] has the operator form $(S_0^+ S_k^- + S_0^- S_k^+)$. Z-rotation in the Pauli space of the atom, performed by the unitary operator $\exp(-i\varphi S_0^z)$, produces the phase factors in this interaction term:

$$\exp(i\varphi S_0^z)\,(S_0^+ S_k^- + S_0^- S_k^+)\,\exp(-i\varphi S_0^z) = \exp(i\varphi)\, S_0^+ S_k^- + \exp(-i\varphi)\, S_0^- S_k^+. \qquad (18)$$

At $\varphi = \pi$ the interaction changes its sign. Consequently, one might expect that if such phase shifts are performed repeatedly and with sufficiently fast rate, the interaction between the atom and each of the field oscillators will be effectively averaged to zero. This would decouple the atom from the electromagnetic field and increase the lifetime in the excited state. Different forms of decoupling, as an example, for an atom, driven by a strong field, in a resonance cavity [10], or by coherent excitation of overlapping resonances [11] have been proposed.

In practice, resonant laser pulses, depending on the relative phase, can directly produce only x- and y-rotations in the interaction (rotating) frame. (We should note that directions in the Pauli space are not related to the directions in the real space). Z-rotation by $\varphi$ can be realized by a composite pulse, as consecutive x, y, and – x rotations [12]:

$$\exp[i(\pi/2)S_0^{x}]\, \exp(i\varphi S_0^{y})\, \exp[-i(\pi/2)S_0^{x}] = \exp(i\varphi S_0^{z}). \qquad (19)$$

A single laser pulse can be converted into a composite z-pulse by splitting the beam into three and introducing different delays for the three paths. Additionally, the delays should be fine-tuned to provide $\pi/2$ phase shifts ($\lambda/4$) between the second and the first, and between the third and the second sub-pulses. The first and the third sub-pulses should be $\pi/2$ pulses, while the attenuation of the second sub-pulse can be used to adjust the angle $\varphi$ of the effective z-rotation.

In a simulation, we neglected the duration of the composite z-pulses and assumed that the multi-pulse sequence produces instantaneous phase kicks, following with the repetition time $\tau_r \ll \tau_F$. The simulation shows that the pulse sequence produces absolutely no effect on the excited state population decay. The simplest explanation for this might be that the atom, at any moment in time, is fully described by the populations of its two states (which are not changed by z-rotations), and that there are no correlations between the atom and the radiation field that may be affected by the z-rotations. Such picture also seems to be consistent with the observed exponential decay of the excited state population. However, the multi-pulse train of phase kicks produces a dramatic change of the emitted spectrum. The results are shown in Fig. 6.

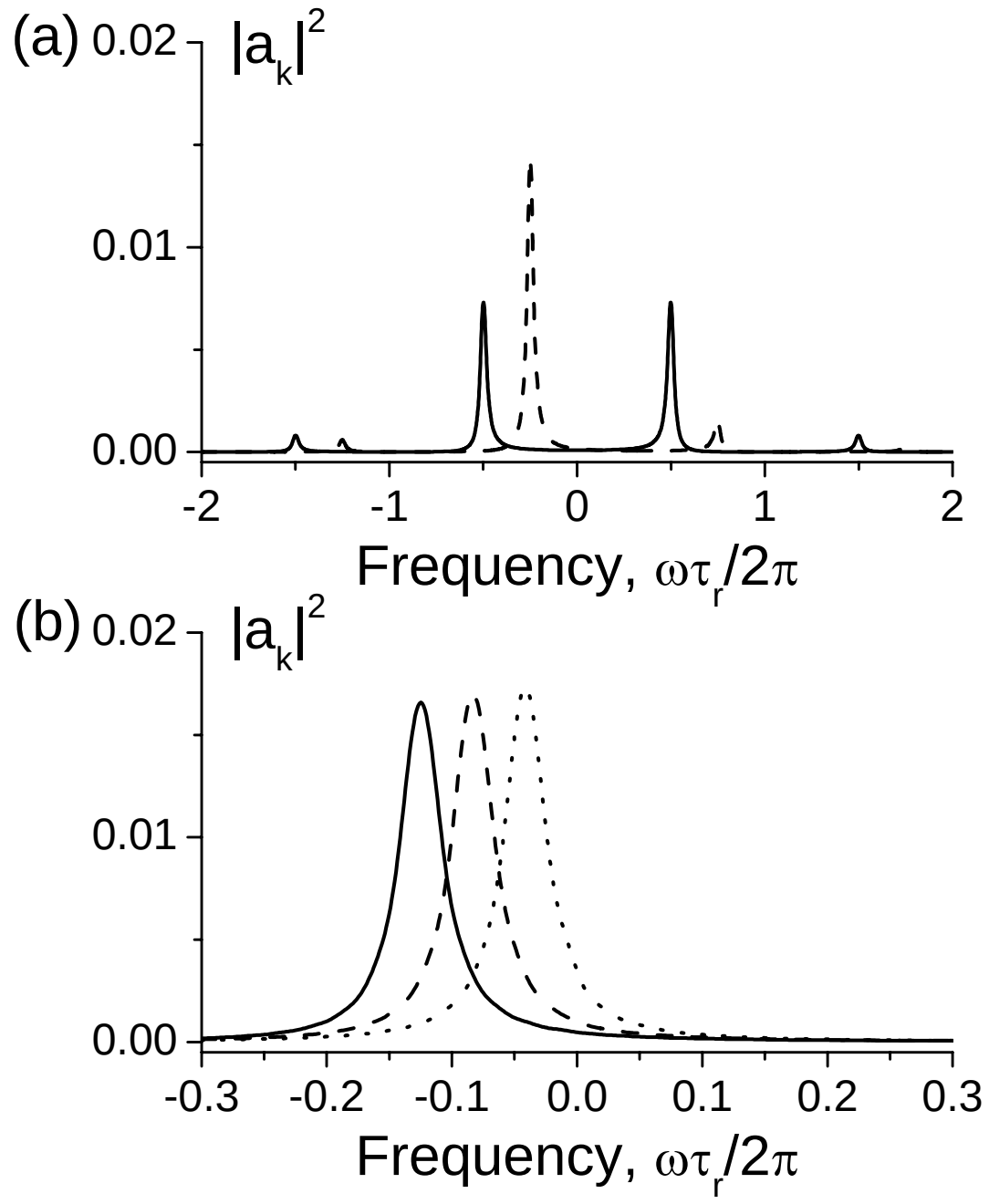

**Fig. 6** The spectrum of spontaneous emission when the atom is irradiated by a multi-pulse sequence of phase kicks. The frequency is in units of the repetition rate $1/\tau_r$, where the interval between z-pulses is $\tau_r = \tau_F/25$. (a) $\varphi = 180^o$ for the solid line and $90^o$ for the dashed line. (b) $\varphi = 45^o$, $30^o$, and $15^o$ respectively for the solid, dashed, and dotted lines.

At $\varphi = \pi$ (solid line in Fig. 6a) the spectrum consists of two peaks at frequencies $\pm\pi/\tau$ and smaller satellites separated by the repetition frequency $2\pi/\tau_r$. Upon decreasing $\varphi$, the total spectral intensity becomes concentrated in the central peak, shifted from the resonance frequency by $\varphi/\tau_r$, which is equal to the average frequency of the phase rotation. It is interesting that this frequency shift is not any integer of the modulation frequency $2\pi/\tau_r$ but can be changed in a continuous way by varying $\varphi$, as can be seen in Fig. 6b. The intensities of the central peak and the satellite peaks are given by the squared Fourier coefficients of the periodic function

$$f(t) = \exp\, \{i\, \varphi \int_0^t dt' \,[\Sigma_n\, \delta(t'\text{-}n\tau_r) - 1/\tau_r]\}. \tag{20}$$

Modification of the spectrum by a sequence of phase kicks suggests that, in the process of spontaneous emission, there exist long-lived phase correlations between the atom and the radiation field. These correlations, quantified as

$$c_{jk} = < S_j^+ S_k^- + S_j^- S_k^+ > = a_j\, a_k^* + a_k\, a_j^* , \tag{21}$$

are presented in Fig. 7 for different frequency offsets of the field oscillator. The correlations are shown for the case when the atom is unperturbed by the pulse sequence, and calculated for $10^6$ field oscillators using the exact solution in Section 3. One can notice that the correlation is surprisingly strong even for the oscillators with frequencies well outside the central part of the emission spectrum.

The field oscillators also remain strongly correlated between themselves. These correlations are shown in Fig. 8. At $t >> \tau_F$, the absolute values of the correlations reach a stationary value $|c_{jk}|/(|a_j||a_k|) = 2$. Therefore, after a photon is emitted, the state of the entire system cannot be fully described by probabilities and doesn't have a simple classical interpretation.

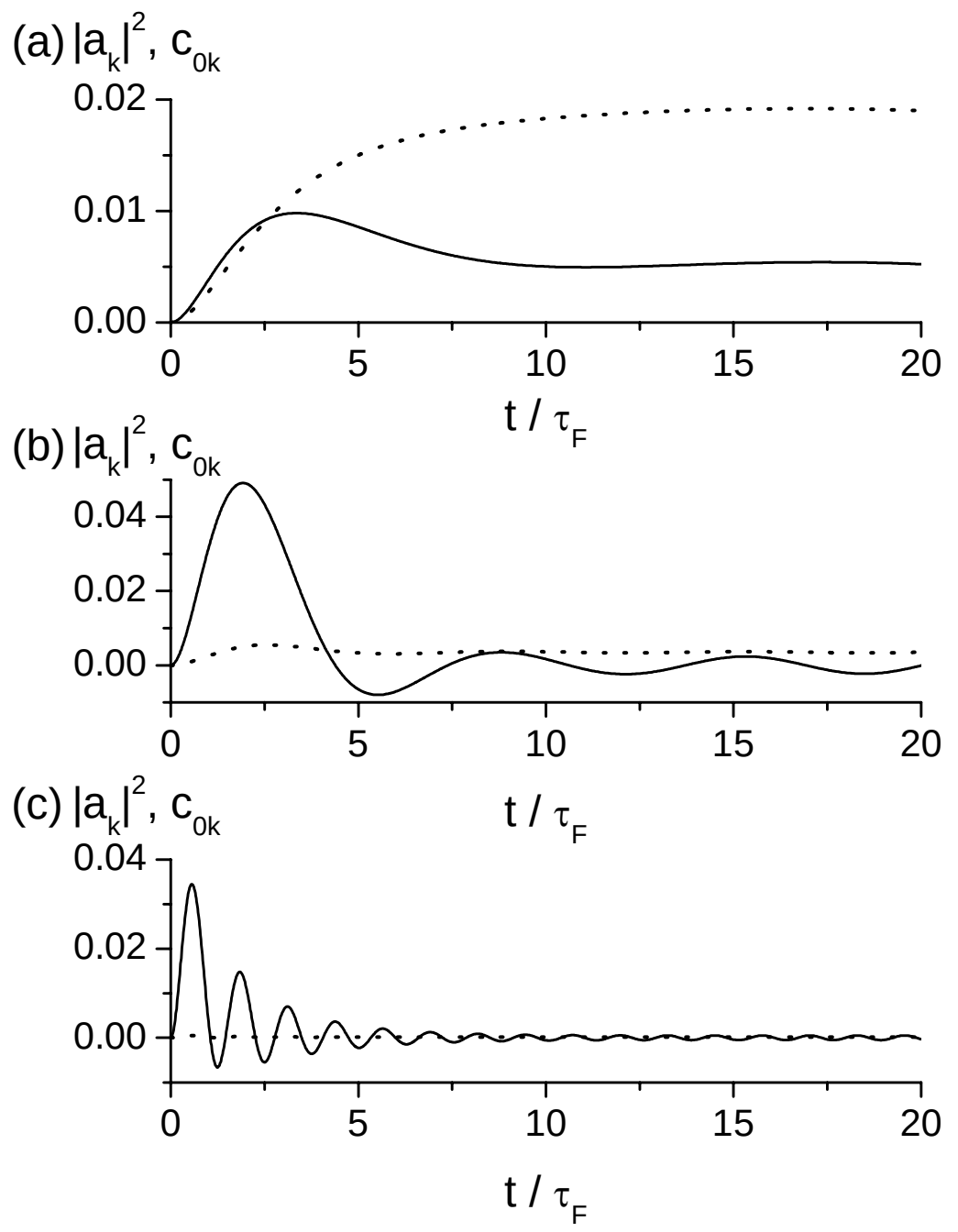

**Fig. 7** Time dependence of correlations $c_{0k}$ (solid) and populations $|a_k|^2$ (dotted) for the oscillators with the frequency offsets (a) $\omega\tau_F = -0.1$, (b) $\omega\tau_F = -1.0$, and (c) $\omega\tau_F = -5.0$.

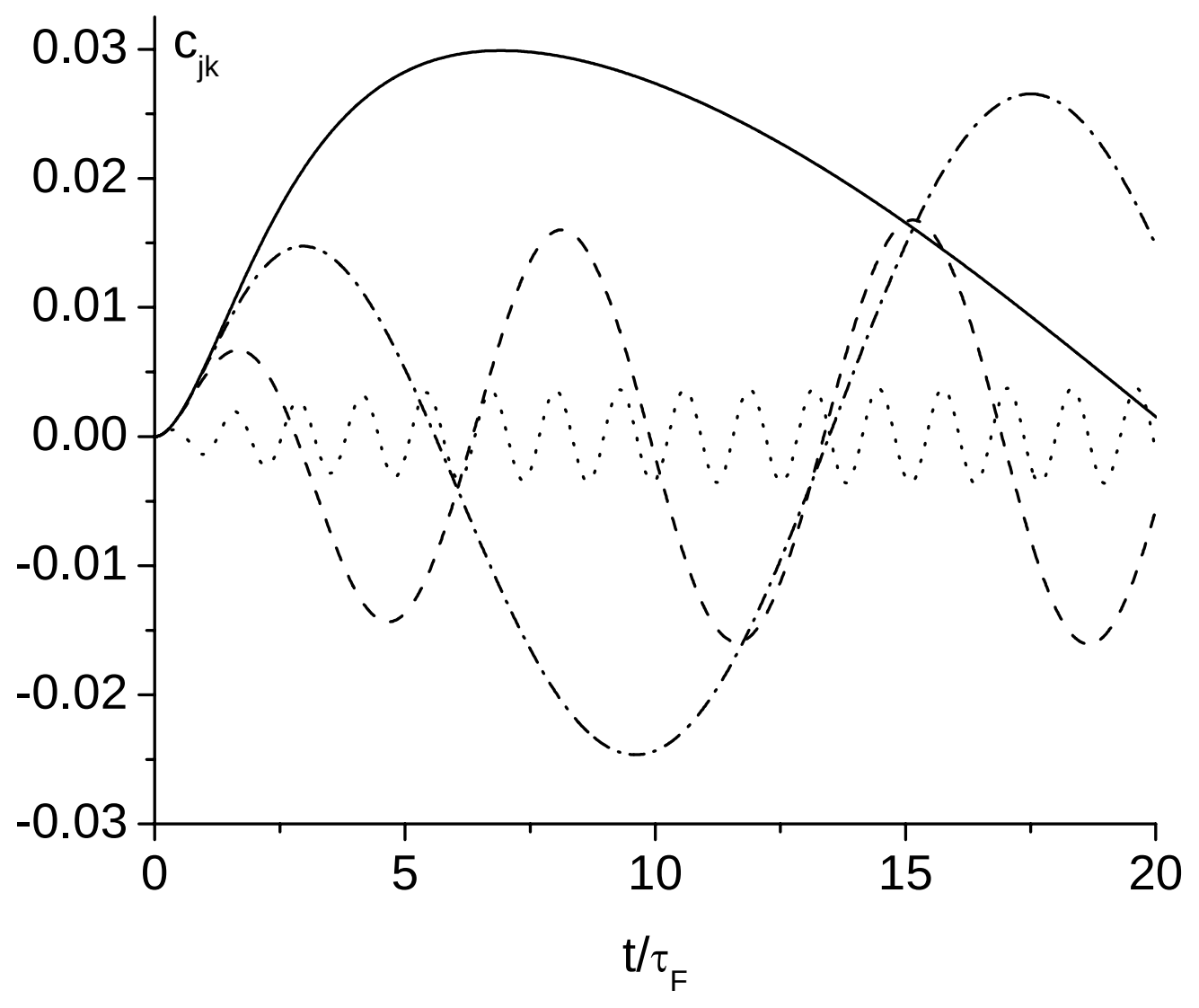

**Fig. 8** Time dependence of correlations $c_{jk}$ for the oscillators with $\omega_j\tau_F = -0.1$ and $\omega_k\tau_F = -0.2$ (solid), -0.5 (dash-dot), -1.0 (dashed), and -5.0 (dotted).

## 5. EMISSION BY A PAIR OF ATOMS. A PERTURBED SYMMETRY

In this section, we present the results for spontaneous emission by a pair of atoms. It is supposed that the atoms are at a very short distance from one another (much smaller than the wavelength), so that the coupling constants between the atom and field oscillators are the same for the two atoms. Dicke analyzed this problem [13] by assuming that a compact multi-atom system is coupled to the radiation field by its total dipole moment. Then, a symmetry-based approach has been used to introduce "super-radiant" and "non-radiant" states of the system. The phenomenon of super-radiant emission by a two-atom system has been observed experimentally [14]. The exact solution for a multi-atom system coupled to a single radiation mode is given in [15].

Let us denote the states with the excitation on the first or on the second atom, with the field oscillators in their ground states, as $|10\rangle$ and $|01\rangle$, respectively. The inclusion of a second atom adds only one state to the single-photon subspace $\{|\Psi_k\rangle\}$. The Hamiltonian (3) is modified by addition of one more "cross" of the interaction constants. Again, we will be using a pseudo-1D model with equal coupling constants. According to [13], the symmetric ("triplet") linear combination $|t\rangle = 2^{-1/2}(|10\rangle + |01\rangle)$ is a fast decaying "super-radiant" state, while the anti-symmetric ("singlet") combination $|s\rangle = 2^{-1/2}(|10\rangle - |01\rangle)$ is a "non-radiant" state with infinite lifetime. One can verify directly that the state $|s\rangle$ is an eigenstate of the Hamiltonian. Our numerical simulation confirmed this prediction. For the initial state $|10\rangle$ with the first atom excited, the excited state population of the first atom exponentially decays to a stationary value of ¼. Simultaneously, the population of the excited state for the second atom increases to the same stationary value of ¼. The emitted spectrum is centered at the atoms' resonance frequency, and the linewidth is doubled compared to the emission by a single atom. This behavior is consistent with viewing the initial state as a sum of two states: $|10\rangle = 2^{-1/2}(|t\rangle + |s\rangle)$, where one of the states, $|t\rangle$, has a doubled decay rate, while the other state, $|s\rangle$, is stationary.

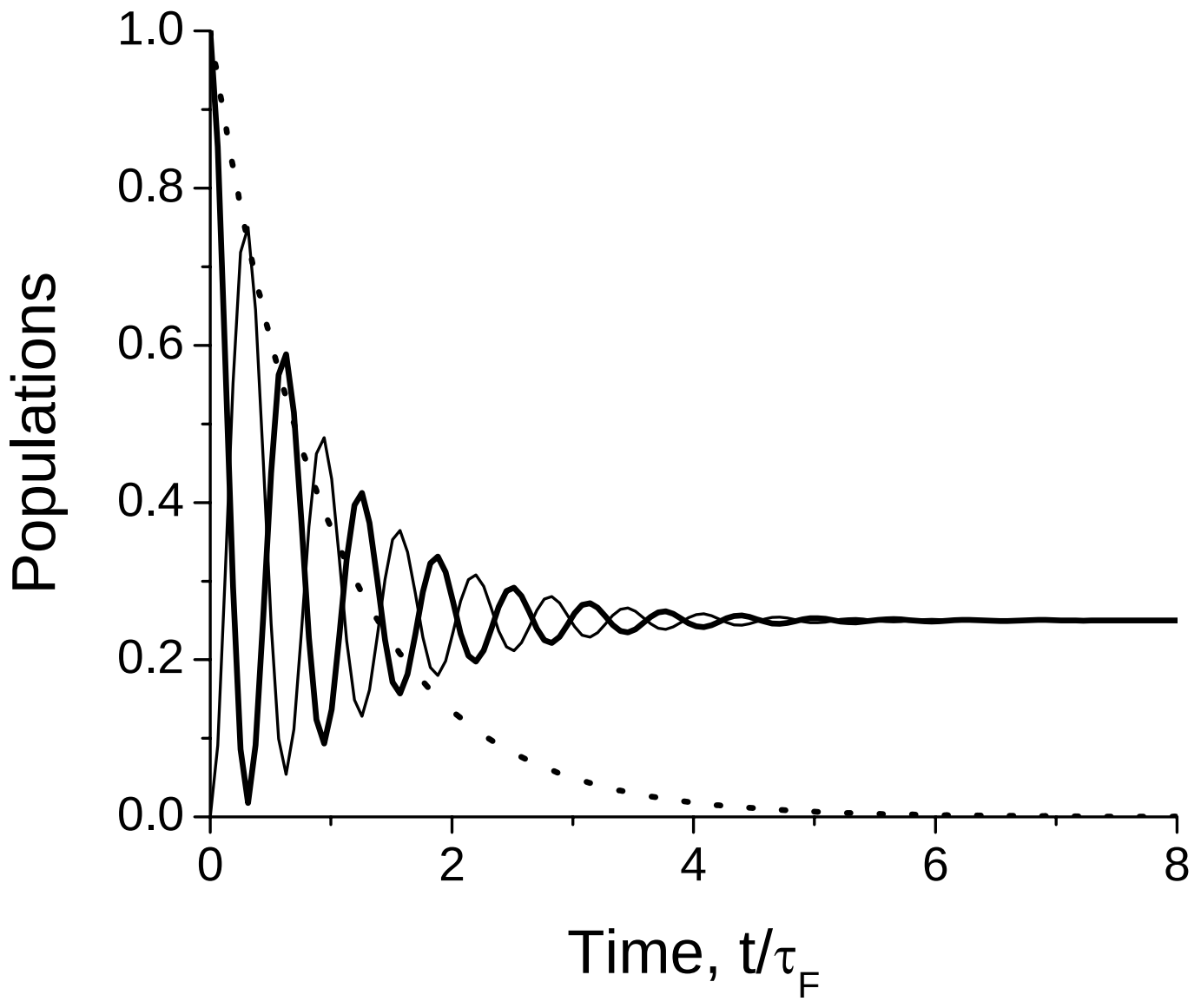

**Fig. 9** Populations of the atomic excited states at dipolar coupling $\omega_d = 5 \times 2\pi\eta^2/\varepsilon$. The resonance frequencies of the two atoms are equal. Thick solid line: population of the state with the first atom excited; thin solid line: population of the state with the second atom excited; dotted line: population decay for a single atom.

Two atoms in a close proximity experience a direct dipole-dipole interaction with one another [16,17]. As an example, two hydrogen atoms at 10 nm, which is about one-tenth of the wavelength of the hydrogen $2p_z \rightarrow 1s$ transition (~121.6 nm), will have a direct dipole-dipole interaction five times stronger (in frequency units) than the linewidth of the hydrogen spontaneous emission spectrum for this transition. Inclusion of the dipole-dipole inter-atomic interaction in the simulation shows (Fig. 9) that the interaction causes a fast exchange of the atomic populations. At the same time, the stationary populations of ¼ do not change, as a consequence of the fact that the dipole-dipole interaction does not spoil the symmetry, and the anti-symmetric state $|s\rangle$ is still an eigenfunction of the Hamiltonian. The corresponding emitted spectrum is shown in Fig. 10. The line is shifted by the dipolar coupling, and its width is doubled compared to the emission by a single atom. In this simulation, it was assumed that the interatomic vector is perpendicular to the atomic dipole moments. The latter determined the sign of the dipolar coupling and the sign of the corresponding frequency shift in the spectrum.

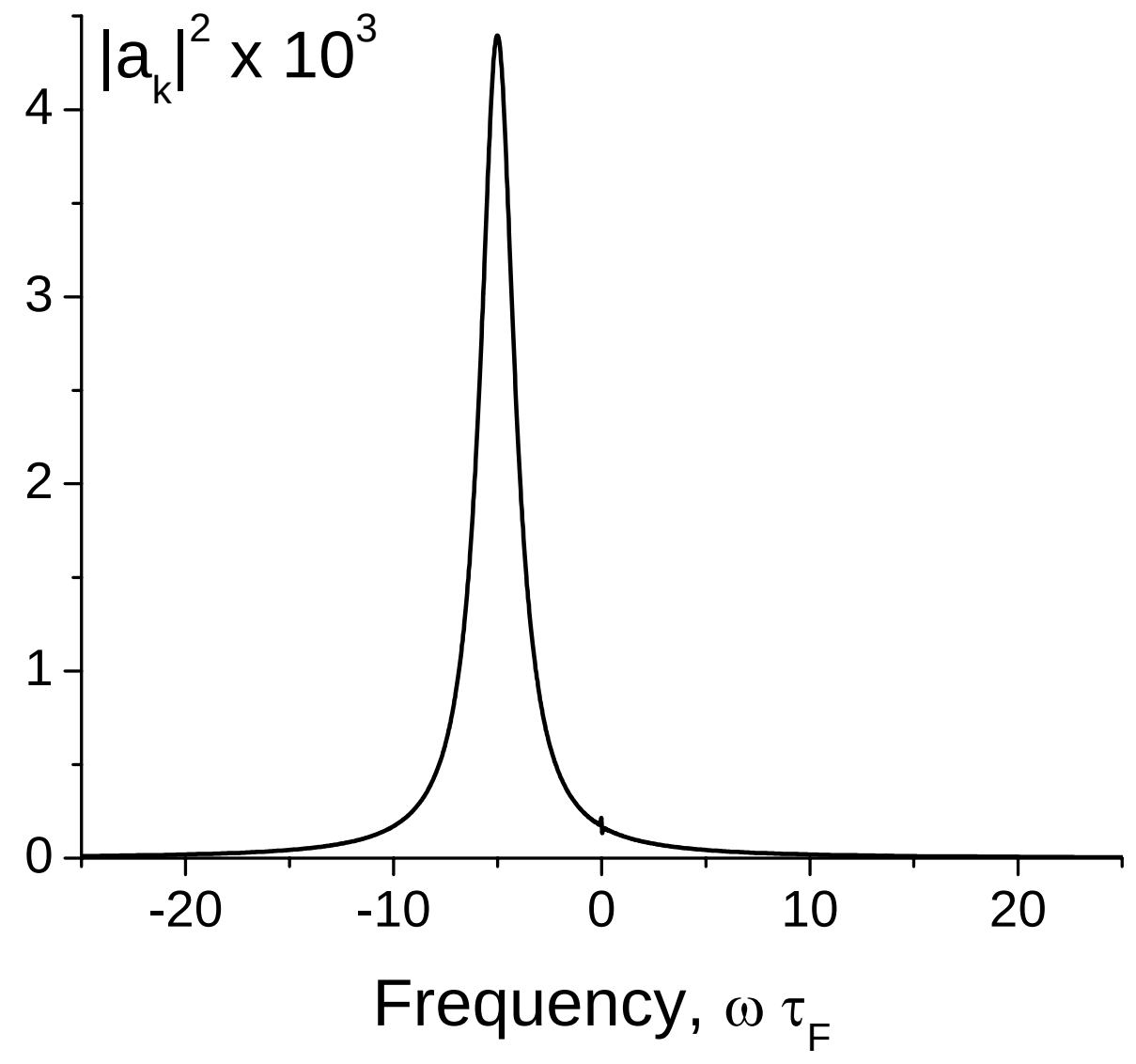

**Fig. 10** The spectrum emitted at $t = 8\tau_F$ for the dynamics shown in Fig. 8.

As a consequence of the system's symmetry, the anti-symmetric state $|s\rangle$ remains uncoupled from the electromagnetic field and does not contribute to the spectrum. Similar to the NMR experiments [18], where symmetry breaking has been used to access the long-lived singlet states, one may hope that a distortion of the symmetry in the two-atom system will result in the light emission by the state $|s\rangle$. The symmetry can be perturbed by a difference in the resonance frequencies of the two atoms. When this difference is smaller than the dipole-dipole coupling, it is averaged by the dipolar interaction and produces little effect. On the other hand, when it is too large, the two atoms behave as independent uncoupled systems. The most interesting behavior happens when the difference in the resonance frequencies is comparable to the dipolar frequency. The results are shown in Fig. 11 for the case when the resonance frequencies of the two atoms are shifted by $\pm\ \omega_d$. One can see that the spectra, for different initial conditions, contain two peaks, one broad and one narrow. The linewidth of the narrow peak is much less than the natural linewidth $\omega_F = 2\pi/\tau_F$. Relative intensities of the broad and narrow components depend on which of the two atoms has been initially excited [Figs. 11 (a) and (b)]. It is interesting that the spectra for these two initial conditions are practically the same as the ones emitted when the atoms are initially prepared in the superposition states $|s\rangle$ and $|t\rangle$ [Figs. 11 (c) and (d)]. The linewidth of the narrow spectral component can be made arbitrary small by decreasing the difference in resonance frequencies. However, such a decrease also reduces the intensity of the narrow spectral component.

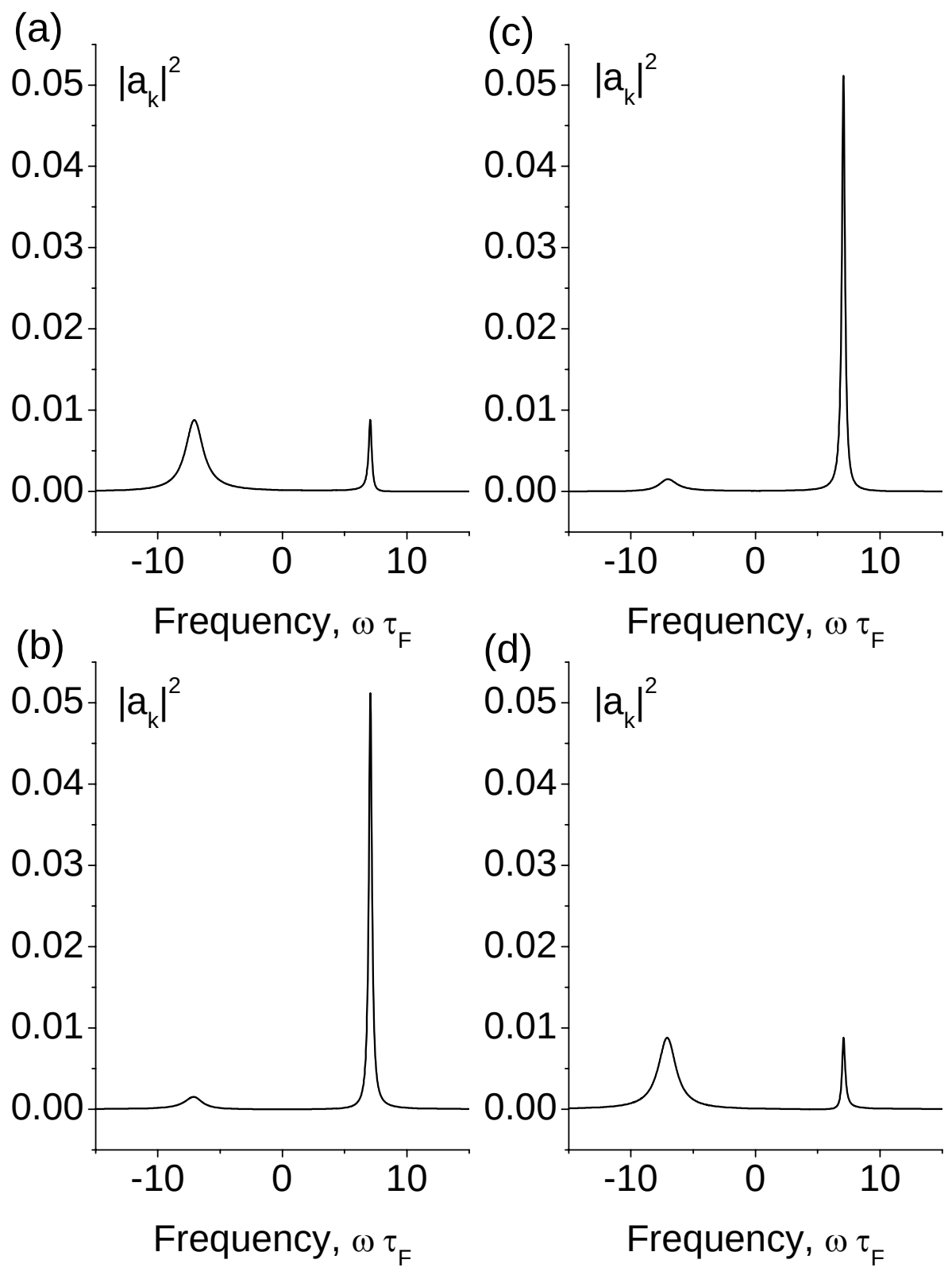

**Fig. 11** The spectra emitted by two atoms with resonance frequencies $\pm\,\omega_d$, where $\omega_d$ is the dipolar coupling constant. The initial conditions are: (a) the atom with resonance frequency $+\omega_d$ is excited; (b) the atom with resonance frequency $-\omega_d$ is excited; (c) the anti-symmetric state $|s\rangle$; (d) the symmetric state $|t\rangle$.

## 6. CONCLUSION

The computational power of modern computers allows direct simulation of the process of spontaneous emission, if the dynamics is limited by a single-photon subspace. With up to 20k field oscillators, the finite discrete model provides results which are very close to the thermodynamic limit, especially in the pseudo-one-dimensional case. Explicit dynamics, obtained as a time-dependent wavefunction of the combined system: atom(s) +

electromagnetic radiation field, reveal many interesting details. A better understanding of this complex collective motion is essential for designing methods of manipulating such dynamics. In this paper, we presented two simple examples demonstrating that both the frequency and the linewidth of the emitted spectrum can be controlled.

Simulations similar to the described above will be helpful in developing new spectroscopic techniques. They can also be used in studying the fundamental process of quantum decoherence. Field oscillators, even within a single-photon subspace, can provide very complex "mixing" dynamics and serve as a thermodynamic bath with explicit quantum-mechanical description. Models of systems with small number of degrees of freedom, coupled to such bath, can be used for elucidating the role of environment in quantum dynamics.

## ACKNOWLEDGEMENTS

The work was supported in part by NSF (JSL and AK), Walsh University (MR), and by an allocation of computing time from the Ohio Supercomputer Center. AK thanks V.A. Atsarkin and P.G. Eliseev who in 1997 attracted his attention to this problem. The authors thank K. A. Khitrin for discussions.